\documentclass[%twocolumn,doublespacing,superscriptaddress,
fleqn]{elsart}
\usepackage{amsmath,amssymb,dcolumn}

\def\be{\begin{equation}}
\def\ee{\end{equation}}
\def\bea{\begin{eqnarray}}
\def\eea{\end{eqnarray}}
\def\ba{\begin{array}}
\def\ea{\end{array}}
\def\nn{\nonumber}

\begin{document}
\begin{frontmatter}

\title{Hawking radiation from $(2+1)$-dimensional BTZ black holes}
\author[1]{Qing-Quan Jiang},
\ead{jiangqq@iopp.ccnu.edu.cn}
\author[2]{Shuang-Qing Wu} and
\ead{sqwu@phy.ccnu.edu.cn}
\author[1]{Xu Cai}
%\ead{xcai@mail.ccnu.edu.cn}
\address[1]{Institute of Particle Physics, Central China Normal University,
Wuhan, Hubei 430079, People's Republic of China}
\address[2]{College of Physical Science and Technology, Central China Normal University,
Wuhan, Hubei 430079, People's Republic of China}
%\date{\today}

\begin{abstract}
Motivated by the Robinson-Wilczek's recent viewpoint that Hawking radiation can be treated as a
compensating energy momentum tensor flux required to cancel gravitational anomaly at the horizon
of a Schwarzschild-type black hole, we investigate Hawking radiation from the rotating $(2+1)$-dimensional
BTZ black hole and the charged $(2+1)$-dimensional BTZ black hole, via cancellation of gauge and
gravitational anomalies at the horizon. To restore gauge invariance and general coordinate covariance
at the quantum level, one must introduce the corresponding gauge current and energy momentum tensor
fluxes to cancel gauge and gravitational anomalies at the horizon. The results show that the values
of these compensating fluxes are exactly equal to those of $(1+1)$-dimensional blackbody radiation
at the Hawking temperature.
\end{abstract}

\end{frontmatter}

\newpage

%%%%%%%%%%%%%%%%%%%%%%%%%%%%%%%%%%%%%%%%%%%%%%%%%%%%%%%%%%%%%
\section{Introduction}\label{intro}
%%%%%%%%%%%%%%%%%%%%%%%%%%%%%%%%%%%%%%%%%%%%%%%%%%%%%%%%%%%%%

Classically, a black hole is black but, according to quantum mechanics, it can emit any kinds of particles
via thermal radiation. This became known as Hawking evaporation soon after Stephen Hawking first discovered
it \cite{SWH}. Heuristically, there exist several derivations of Hawking radiation \cite{JWC,PW}. Among
them, one method is attributed to the trace anomaly in the conformal symmetry \cite{CF}, where the authors
determined the full momentum energy tensor by using symmetry arguments and the conservation law of the energy
momentum tensor together with the trace anomaly. Although their derived result is in quantitative agreement
with that of Hawking's, it is difficult to generalize their method to the cases of higher dimensional black
holes by using partial wave analysis since Hawking flux is connected with the anomaly through an integral over
the whole spacetime. In addition, their observation was based upon the assumptions that the fields were massless
and there was no back-scattering effect. Thus, the method appears to be rather special.

Recently, Robinson and Wilczek \cite{RW} proposed a new derivation of Hawking radiation via gravitational
anomaly at the horizon of a Schwarzschild-type black hole. Here, gravitational anomaly stems from the fact
that one wants to integrate out the horizon-skimming modes, whose contributions would give a divergent energy
momentum tensor because these modes make the time coordinate ill-defined at the horizon. Thus, the effective
theory formed outside the black hole horizon to exclude the offending modes becomes chiral near the horizon,
and contains an anomaly with respect to general coordinate symmetry. To restore general coordinate covariance
at the quantum level, one must introduce an energy momentum tensor flux to cancel gravitational anomaly at
the horizon. The result shows that the compensating energy momentum tensor flux has an equivalent form to
that of Hawking radiation. A key in their derivation is to introduce a dimensional reduction technique that
generates the physics near the horizon of the original higher dimensional black holes, which can be effectively
described by an infinite collection of two-dimensional fields. In the simplest case for a scalar field,
gravitational anomaly can only occur in theories with chiral matter coupled to gravity in a spacetime of
dimensions $4n+2$, where $n$ is an integer. Obviously the derivation of Hawking radiation via the anomalous
point of view is only dependent on the anomaly at the horizon, so the method may be more universal. Following
this method and considering gauge and gravitational anomalies at the horizon, Iso \textit{et al}. \cite{IUW1}
have investigated Hawking radiation of charged particles from Reissner-Nordstr\"{o}m black holes. The
result shows that the electric current and energy momentum tensor fluxes, required to cancel gauge and
gravitational anomalies at the horizon, are exactly equal to those of Hawking radiation. Subsequently,
there have been considerable efforts to generalize this work to other cases, and all the obtained results
are very successful in supporting the Robinson-Wilczek's prescription \cite{IUW2,MV,recent}.

On the other hand, the $(2+1)$-dimensional Banados-Teitelboim-Zanelli (BTZ) black holes have recently received
a lot of attention for various reasons. This is mainly because in $(3+1)$ or even higher dimensional gravity,
the black hole's properties at the quantum level remain, until now, as a mystery; it is believed that the black
holes in $(2+1)$-dimensions will provide a good relatively simple laboratory and a better understanding for
analyzing general aspects of black hole physics. Another major motivation is due to the AdS/CFT correspondence
\cite{GKP}, which relates thermal properties of black holes in AdS space to that of a dual CFT. Combined with
these reasons, it has attracted much attention in recent years to study the thermal properties of
$(2+1)$-dimensional black holes, especially in the background geometry with a non-vanishing cosmological constant.
In this Letter, we will focus on studying Hawking radiation of $(2+1)$-dimensional BTZ black holes via anomalies
at the black hole horizon. In fact, since the metric of the rotating black holes has an azimuthal symmetry,
the angular momentum is conserved under gauge coordinate transformation. After a dimensional reduction
technique at the horizon, each partial wave of the quantum fields in the black hole can be effectively described
by a two-dimensional charged field with a $U(1)$ gauge charge $m$, where $m$ is an azimuthal quantum number.
One then adopts the same procedure as that for the charged black holes to obtain Hawking radiation of the
rotating black holes via gauge and gravitational anomalies. However, in the dragging coordinate system, the
matter field in the ergosphere near the horizon must be dragged by the gravitational field with an azimuthal
angular momentum because there exists a frame dragging effect in the rotating spacetime, so an observer
at rest in the dragging coordinate system will not find a $U(1)$ gauge flux since the $U(1)$ gauge symmetry
is no longer incorporated in each partial wave for the two-dimensional theory due to the dragging effect.
As far as the charged black holes are concerned, the effective two-dimensional theory exhibits a gauge symmetry
with respect to the Maxwell field of the original black holes. In such a case, if the effective field theory
is formulated outside the black hole horizon to exclude the offending modes at the horizon, it contains an
anomaly with the gauge symmetry.

In this Letter, we are interested in studying Hawking radiation from the BTZ black holes via gauge and
gravitational anomalies at the horizon. Our result supports the Robinson-Wilczek's viewpoint, and shows
that Hawking radiation can be derived from the anomalous cancellation condition and the regularity
requirement at the horizon. For the rotating BTZ black hole in the dragging coordinate system, the
compensating flux of the energy momentum tensor required to cancel the gravitational anomaly at the horizon
is exactly equal to that of Hawking radiation whose spectrum is given by the Planckian distribution
without a chemical potential for an azimuthal angular momentum. If restored in the original coordinates,
the $U(1)$ gauge current flux is present to cancel the $U(1)$ gauge anomaly at the horizon. The result shows
that the compensating flux corresponds to the angular momentum flux of Hawking radiation, where the Hawking
distribution contains a chemical potential for an azimuthal angular momentum. In the case of the charged
BTZ black hole, the effective two-dimensional quantum field near the horizon is endowed with the gauge and
general coordinate symmetries, where the gauge symmetry is originated from the electric field of the
charged black hole. In order to restore gauge invariance and general coordinate covariance at the quantum
level, the gauge charge current and energy momentum tensor fluxes are required to cancel gauge and gravitational
anomalies at the horizon. It is shown that the compensating fluxes have an equivalent form to those of
Hawking radiation with an electric chemical potential.

Our Letter is outlined as follows. In Sec. \ref{sbh}, we review (slightly different from \cite{RW}) the
Robinson-Wilczek's method in the case of a Schwarzschild-type black hole. Adopting the dragging coordinate
system, Sec. \ref{rbh} is devoted to proving that the energy momentum tensor flux, derived by cancelling
the anomaly with respect to general coordinate covariance at the horizon of the rotating BTZ black hole,
is equal to that of Hawking radiation. At the same time, we also point out that the $U(1)$ gauge current
flux, which occurs in the original coordinates, corresponds to the angular momentum flux of Hawking radiation
with a chemical potential for an azimuthal angular momentum. In Sec. \ref{cbh}, we show that gauge and
gravitational anomalies at the horizon of the charged BTZ black hole can be cancelled by the electric
and energy momentum tensor fluxes of Hawking radiation. Finally, Sec. \ref{core} ends up with our brief
discussions.

%%%%%%%%%%%%%%%%%%%%%%%%%%%%%%%%%%%%%%%%%%%%%%%%%%%%%%%%%%%%%
\section{Hawking radiation from a Schwarzschild-type black hole}\label{sbh}
%%%%%%%%%%%%%%%%%%%%%%%%%%%%%%%%%%%%%%%%%%%%%%%%%%%%%%%%%%%%%

The metric of a $d$-dimensional Schwarzschild-type black hole takes the form
\be
ds^2 = f(r)dt^2 -\frac{dr^2}{f(r)} -r^2d\Omega^2_{(d-2)} \, ,
\ee
where, $f(r)$ is dependent on the matter distribution. The horizon $(r = r_H)$ is determined by $f(r_H)
= 0$. The surface gravity at the horizon is given by $\kappa = (1/2)f_{,r}(r_H)$. Near the horizon, after
performing the partial wave decomposition in terms of the spherical harmonics and then transforming to
the tortoise coordinate defined by $dr_*/dr\equiv f^{-1}(r)$, the effective radial potential vanishes
near the horizon. Now, the quantum field in the $d$-dimensional black holes can be effectively described
by an infinite collection of quantum fields in the ($1+1$)-dimensional spacetime
\be
ds^2 = f(r)dt^2 - \frac{dr^2}{f(r)} \, ,
\ee
together with a background dilaton field $\Psi = r^{(D-2)}$, whose effect on the flow of the energy momentum
tensor is dropped since the involved background is static.

It is well-known that an anomaly will occur when the symmetry or the corresponding conservation law, which
is valid in the classical theory, is violated in the quantized version. In the static Schwarzschild-type
black hole, there is a global Killing vector, which generates a general coordinate symmetry of the spacetime.
If we formulate the effective field theory outside the horizon to integrate out the classically irrelevant
ingoing modes at the horizon, the effective theory becomes chiral since there is a divergent energy momentum tensor
due to a pile up of the offending modes, and it contains a gravitational anomaly with respect to the general
coordinate symmetry, which takes the form of the nonconservation of the energy momentum tensor. Taking the
simplest case for a scalar field, a gravitational anomaly can only occur in theories with chiral matter coupled
to gravity in spacetimes of dimensions $4n+2$, where $n$ is an integer. In $1+1$ dimensions, the consistent
anomaly for the energy momentum tensor reads \cite{LE}
\be
\nabla_\mu T_\nu^\mu \equiv \frac{1}{\sqrt{-g}}\partial_\mu N^\mu_\nu
= \frac{1}{96\pi\sqrt{-g}}\epsilon^{\beta\mu}
\partial_{\mu}\partial_{\alpha}\Gamma_{\nu~\beta}^{~\alpha} \, ,
\ee
where $\epsilon^{01} = 1$. To restore general coordinate covariance at the quantum level, one must introduce
a compensating flux of the energy momentum tensor whose contributions exactly cancel the gravitational anomaly at the
horizon. We expect that the compensating flux of the energy momentum tensor is precisely equal to that of Hawking
radiation, so as to guarantee the general covariance of the effective theory at the quantum level.

Now, we will specifically determine the flux of the energy momentum tensor. This flux is required to cancel
gravitational anomaly at the horizon, so we expect that it is exactly equal to that of Hawking radiation.
Outside the horizon of the black hole, the effective field theory is formulated to exclude the horizon-skimming
modes. We split the region outside the horizon into two patches: Near the horizon $ r_H \leq r \leq r_H +\varepsilon$,
when excluding the ingoing modes, the effective theory is chiral here, and contains a gravitational anomaly.
The nonconservation of the energy momentum tensor in this region specifically satisfies the anomalous equation
as $\partial_r T^r_{(H)t} = \partial_r N_t^r(r)$, where
\be
N^r_t(r) = \frac{1}{192\pi}{(f_{,r}^2 +f_{,rr}f)}.
\ee
In the other region $r_H +\varepsilon \leq r$, since the number of the ingoing and outgoing modes is always
identical, there is no anomaly with respect to general coordinate symmetry, and the energy momentum tensor
should be conserved there, which means $\partial_r{T^r_{(o)t}} = 0$. In the classical theory, general
coordinate covariance of the classical action tells us that its variation must satisfy $-\delta_\lambda
W = \int d^2x \sqrt{-g_{(2)}} \lambda^\nu \nabla_\mu T_\nu^\mu = 0$, where $\lambda^\nu$ is a variation
parameter. The variation of the effective action, under a general coordinate transformation, can be read off
\bea
-\delta_\lambda W &=& \int d^2x \lambda^t \partial_r T_t^r \nn \\
&=& \int d^2x \lambda^t \Big\{\partial_r \big[N_t^r(r)H(r)\big] \nn \\
&& +\big[T_{(o)t}^r -T_{(H)t}^r +N_t^r(r)\big]\delta \big(r -r_H -\epsilon\big)\Big\} \, ,
\eea
where the energy momentum tensor combines contributions from the two regions outside the horizon; that is,
$T_t^r = T_{(o)t}^r\Theta_+(r) +T_{(H)t}^rH(r)$, in which $\Theta_+(r) = \Theta(r -r_H -\epsilon)$ is the
scalar step function and $H(r) = 1 -\Theta_+(r)$ is a scalar ``top hat'' function. Obviously, we have not
taken the quantum effect of the ingoing modes into account here. If it is incorporated, the first item
should be cancelled since its contribution to the energy momentum tensor is $-N_t^r(r)H(r)$. In order to
restore general coordinate covariance at the quantum level, the coefficient of the delta-function should
vanish, which yields
\be
a_o = a_H -N_t^r(r_H) \, ,
\ee
where $a_o = T^r_{(o)t}$ is an integration constant and $a_H = T_{(H)t}^r +N_t^r(r_H) -N_t^r(r)$ is the
value of the energy flow at the horizon. To ensure the regularity of the physical quantities at the horizon,
the covariant energy momentum tensor should vanish at the horizon, which means
\be
\tilde{T}_{(H)t}^r = T_{(H)t}^r +\frac{1}{192\pi}\Big(f{f_{,rr}} -2f_{,r}^2\Big) = 0 \, .
\label{temp}
\ee
The validity of this condition has been clarified in \cite{IUW2}. Thus we have
\be
a_H = 2N_t^r(r_H) = \frac{\kappa^2}{24\pi} \, ,
\ee
where $\kappa = f_{,r}(r_H)/2$ is the surface gravity at the black hole horizon. The total flux of the energy
momentum tensor is
\be
a_o = N_t^r(r_H) = \frac{\pi}{12}T_H^2 \, ,
\ee
where $T_H = \kappa/(2\pi) = f_{,r}(r_H)/(4\pi)$ is the Hawking temperature of the black hole.

Next, we shall prove the total compensating flux of the energy momentum tensor is exactly equal to that of
Hawking radiation. For fermions, the Hawking radiant spectrum of a Schwarzschild-type black hole is given
by the Planckian distribution $N(\omega) = 1/[\exp(\frac{\omega}{T_H}) +1]$, so the energy momentum
tensor flux of Hawking radiation from the black hole is
\be
F_{H} = \int_0^\infty \frac{\omega}{\pi}N(\omega)d\omega = \frac{\pi}{12}T_H^2 \, .
\ee

Obviously, the flux of the energy momentum tensor, required to cancel gravitational anomaly at the horizon
and to restore general coordinate covariance at the quantum level, is precisely equal to that of Hawking
radiation. This shows that Hawking radiation can also be derived from the conditions of anomaly cancellation
and the regularity requirement at the horizon of the black hole. To verify its validity, Iso \textit{et al}.
have subsequently studied Hawking radiation from the Kerr and Kerr-Newman black holes via gauge and
gravitational anomalies. The stationary rotating black holes are normally characterized by the angular
momentum and energy momentum tensor conservation. If we have omitted the classically irrelevant ingoing modes
at the horizon, these physical quantities become anomalous because gauge and general coordinate symmetries
are destroyed. To restore these symmetries, one must introduce gauge current and energy momentum tensor fluxes
to cancel gauge and gravitational anomalies at the horizon. These compensating fluxes can be shown to be exactly
equal to those of Hawking radiation. However, provided that a free falling observer views the situation from the dragging
coordinate system of the rotating black holes, he would not, upon passing the horizon, find a gauge flux since
the dragging coordinate system behaves like a locally non-rotating coordinate system, showing that the effective
theory does not possess the $U(1)$ gauge symmetry originating from the axisymmetry of the rotating black holes. In the
subsequent section, we will extend the Robinson-Wilczek's analysis to the case of the rotating BTZ black hole in the dragging
coordinate system. For a holographic description on gravitational anomalies of the ($2+1$)-dimensional BTZ black
holes see \cite{HA}, for example.

%%%%%%%%%%%%%%%%%%%%%%%%%%%%%%%%%%%%%%%%%%%%%%%%%%%%%%%%%%%%%
\section{Hawking radiation from a rotating BTZ black hole}\label{rbh}
%%%%%%%%%%%%%%%%%%%%%%%%%%%%%%%%%%%%%%%%%%%%%%%%%%%%%%%%%%%%%

The rotating BTZ black hole is an exact solution to the Einstein field equation in a ($2+1$)-dimensional gravity
theory
\be
S=\int dx^3 \sqrt{-g}(^{(3)}R-2\Lambda),
\ee
with a negative cosmological constant $\Lambda = -1/l^2$. Its explicit expression takes the form \cite{BTZ}
\be
ds^2 = f(r)dt^2 -\frac{dr^2}{f(r)} -r^2\Big(d\phi -\frac{J}{2r^2}dt\Big)^2 \, ,\label{eq}
\ee
where the lapse function is
\be
f(r) = -M +\frac{r^2}{l^2} +\frac{J^2}{4r^2},
\ee
$M$ and $J$ are the ADM mass and angular momentum of the rotating BTZ black hole, respectively. The outer
horizon $(r_+)$ and the inner horizon $(r_-)$ are given by the equation $f(r) = 0$. The surface gravity
and the angular momentum of the outer horizon are easily evaluated as
\bea
&&\kappa = \frac{1}{2}\frac{\partial f}{\partial r}\Big|_{r = r_+} = \frac{r_+^2 -r_-^2}{r_+l^2}, \nn\\
&&\Omega_+ = \frac{J}{2r^2}\Big|_{r=r_+ }=\frac{r_-}{r_+l},
\eea
respectively. The coordinate system in Eq. (\ref{eq}) is described by the observer at infinity. If
transforms to the dragging coordinate system, defined by
\be
\psi = \phi -\frac{J}{2r^2}t \, , \qquad \xi = t \, ,
\ee
and then performs a dimensional reduction technique as in Sec. \ref{sbh} on the action of the scalar field
in the ($2+1$)-dimensional rotating BTZ black hole in the dragging coordinate system, one would find that
the physics near the horizon can be effectively described by an infinite collection of two-dimensional scalar
fields on the metric
\be
ds^2 = f(r)d\xi^2 -\frac{dr^2}{f(r)} \, ,
\ee
and a dilaton background $\Psi = r$, whose contributions to the anomalous flux are dropped. Now, we will
study the compensating energy momentum tensor flux and gravitational anomaly at the horizon. In our cases,
the rotating $(2+1)$-dimensional BTZ black hole is at rest in the dragging coordinate system, so only general
coordinate symmetry exists in the two-dimensional reduction. When neglecting the offending modes at the
horizon, the effective action exhibits an anomaly with respect to general coordinate symmetry. To demand
general coordinate covariance at the quantum level, one must introduce an energy momentum tensor flux to
cancel the gravitational anomaly. The effective field theory is formulated outside the horizon to omit
the horizon-skimming modes: In the region $r_+ +\epsilon \leq r$, there is no anomaly, and the energy momentum
tensor satisfies the conservation equation $\partial_rT_{(o)t}^r = 0$, but in the near-horizon region $r_+
\leq r \leq r_+ +\epsilon$, the effective theory contains gravitational anomaly since the offending modes
are integrated out. In terms of the energy momentum tensor, it can be expressed as $\partial_r {T^r_{(H)t}}
= \partial_r {N_t^r(r)}$, where $N^r_t(r) = (f_{,r}^2 +f_{,rr}f)/(192\pi)$. Using the formalism in the
preceding section and demanding general coordinate covariance at the quantum level to hold in the effective
action, one has
\be
c_o = c_H -N_t^r(r_+) = c_H -\frac{1}{192\pi}f_{,r}^2\big|_{r = r_+} \, ,
\ee
where $c_o$ is an integration constant, and $c_H = T_{(H)t}^r +N_t^r(r_+ )-N_t^r(r)$ is the value of the
energy flow at the horizon. Imposing a condition that the covariant energy momentum tensor vanishes at
the horizon [see Eq. (\ref{temp})], the total flux of the energy momentum tensor is
\be
c_o = N_t^r(r_+) = \frac{\pi}{12}T_+^2 \, ,
\label{femt}
\ee
where
\be
T_+ = \frac{\kappa}{2\pi} = \frac{r_+^2 -r_-^2}{2\pi r_+l^2},
\ee
is the Hawking temperature of the black hole.

Thus, one must introduce an energy momentum tensor flux whose value is expressed by Eq. (\ref{femt}) to cancel
the gravitational anomaly and to restore general coordinate covariance for the effective theory. If we take
the quantum effect of the classically irrelevant ingoing modes into account, the introduced energy momentum
tensor flux should be equal to that of Hawking radiation. In the dragging coordinate system, the Hawking
distribution for fermions in the rotating BTZ black hole is given by the distribution $N(\omega) =
1/[\exp(\frac{\omega}{T_+}) +1]$, where the energy $\omega$ is carried by the observer in the dragging
coordinates which is related by the formula  $\omega=\omega'-m\Omega_+$ to the energy $\omega'$ measured
at infinity ($m$ is an azimuthal angular quantum number), so the energy momentum tensor flux of Hawking
radiation reads as
\be
F_{H} = \int_0^\infty\frac{\omega}{\pi}N(\omega)d\omega = \frac{\pi}{12}T_+^2 \, .
\label{fhr}
\ee
Obviously, the right hand sides of Eqs. (\ref{femt}) and (\ref{fhr}) coincide with each other, which shows
that Hawking radiation can be determined from the anomalous point of view. Here, we take as an example the
rotating BTZ black hole in the dragging coordinate system. In fact, after transforming the dragging coordinates
$(\xi, \psi)$ into the original coordinates $(t,\phi)$, one can discover the same result that Hawking
radiation can cancel gauge and gravitational anomalies at the horizon and restore gauge invariance and
general coordinate covariance at the quantum level. Here, a gauge anomaly arises from the destruction of the
$U(1)$ gauge symmetry along $\phi$ direction after excluding the ingoing modes at the horizon. To restore
gauge symmetry of the effective theory, a $U(1)$ gauge current flux must be introduced to cancel the $U(1)$
gauge anomaly. At the quantum level, this gauge current flux should be equal to the angular momentum flux
of Hawking radiation. Note that the Hawking distribution for fermions case is now given by the Planckian
distribution with a chemical potential for an azimuthal angular momentum as $N_{\pm m}(\omega') =
1/[\exp(\frac{\omega'\mp m\Omega_+}{T_+}) +1]$, thus the angular momentum and energy momentum tensor
fluxes are, respectively, given by
\bea
F_m &=& m\int_0^\infty\frac{1}{2\pi}\Big[N_m(\omega') -N_{-m}(\omega')\Big]d\omega'
= \frac{m^2 r_-}{2\pi r_+ l}, \nn \\
F_{H} &=& \int_0^\infty\frac{\omega'}{2\pi}\Big[N_m(\omega') +N_{-m}(\omega')\Big]d\omega'
= \frac{m^2r_-^2}{4\pi r_+^2 l^2} +\frac{\pi}{12}T_+^2 \, .
\eea
These fluxes of Hawking radiation are capable of cancelling $U(1)$ gauge and gravitational anomalies
at the black hole horizon. Hence, the compensating flux to cancel $U(1)$ gauge anomaly is exactly equal
to the angular momentum flux of Hawking radiation whose particle energy $\omega'$ is measured at infinity.

%%%%%%%%%%%%%%%%%%%%%%%%%%%%%%%%%%%%%%%%%%%%%%%%%%%%%%%%%%%%%
\section{Hawking radiation from a charged BTZ black hole}\label{cbh}
%%%%%%%%%%%%%%%%%%%%%%%%%%%%%%%%%%%%%%%%%%%%%%%%%%%%%%%%%%%%%

A static, charged BTZ black hole found for the action (see \cite{BTZ}) is written in the following form as
\be
ds^2 = f(r)dr^2 -\frac{dr^2}{f(r)} -r^2d\phi^2 \, ,
\ee
where the lapse function
\be
f(r)= -M +\frac{r^2}{l^2} -{Q^2}\ln\frac{r^2}{l^2},
\ee
$M$ and $Q$ are the ADM mass and total charge of the black hole, respectively. The horizon of the black hole
$(r = r_+)$ is determined by $f(r_+) = 0$, and the surface gravity is given by
\be
\kappa = \frac{1}{2}\partial_rf|_{r = r_+} = \frac{2r_+^3 -l^4Q^2}{2l^2r_+^2},
\ee
the nonvanishing component for the gauge potential of the Maxwell field is given by $\Phi_t(r) = -Q\ln(r/l)$.
It should be stressed that the expressions presented here for the lapse function and the electric potential
obey not only the differential but also integral forms of the first law of ($2+1$)-dimensional BTZ black hole
thermodynamics with a variable cosmological constant, recently discussed in \cite{FL}.

In the case of a $(2+1)$-dimensional charged BTZ black hole, if one performs the partial wave decomposition,
and then transforms to the tortoise coordinate defined by $dr_* = dr/f(r)$, one finds that the physics
near the horizon can be effectively described by an infinite collection of $(1+1)$-dimensional fields,
each propagating in the ($1+1$)-dimensional spacetime given by
\be
ds^2 = f(r)dt^2 -\frac{dr^2}{f(r)} \, ,
\ee
with a dilaton background $\Psi = r$ and a gauge field $\Phi_t(r)$. Now, the effective two-dimensional
theory for each partial wave has gauge and general coordinate symmetries. If the effective theory is
formulated outside the horizon to exclude the ingoing modes at the horizon, it is chiral here
and contains the anomalies with respect to gauge and general coordinate symmetries, normally called
gauge and gravitational anomalies, respectively.

Now, we study the charge flux and gauge anomaly at the horizon. As the effective two-dimensional chiral
theory contains gauge anomaly at the horizon, a gauge current flux must be introduced to restore gauge
invariance. At the quantum level, we expect that the compensating flux is equal to the charge flux of
Hawking radiation. The effective field theory is still formulated outside the horizon to exclude the
horizon-skimming modes. If we divide the region outside the horizon into two parts: In the region $r_+
\leq r \leq r_+ +\epsilon$, when omitting the classically irrelevant ingoing modes, the effective theory
becomes chiral here, the gauge current exhibits an anomaly, and satisfies the anomalous equation,
$\partial_rJ_{(H)}^r = e^2\partial_r \Phi_t(r)/(4\pi)$; In the other region $r_+ +\epsilon \leq r$,
there is no anomaly in the gauge symmetry, so the gauge current is conserved there, and is determined by
$\partial_rJ_{(o)}^r = 0$. Applying the analysis in Sec. \ref{sbh}, the variation of the effective
action, under gauge transformations, satisfies
\be
-\delta_\lambda W = \int dtdr\lambda\Big\{\partial_r \big[\frac{e^2}{4\pi}\Phi_t(r)H(r)\big]
+\big[J_{(o)}^r -J_{(H)}^r +\frac{e^2}{4\pi}\Phi_t(r)\big]\delta(r -r_+ -\epsilon)\Big\} \, ,
\ee
where we have not taken into account the quantum effect of the ingoing modes. If one elevates the effective
theory to the quantum level, the first term should be cancelled by the quantum effect of the classically
irrelevant ingoing modes since its contribution to the total current is $ -e^2\Phi_t(r)H(r)/(4\pi)$. To
ensure gauge invariance of the total effective action, the coefficient of the delta function should vanish,
that is
\be
D_o = D_H-\frac{e^2}{4\pi}\Phi_t(r_+) \, ,
\ee
where $D_o=J_{(o)}^r $ is an integration constant and $D_{H}=J_{(H)}^r -e^2[\Phi_t(r) -\Phi_t(r_+)]/(4\pi)$
is the value of the consistent gauge current at the horizon. To ensure the regularity requirement at the
horizon, the covariant current formulated by the consistent one as $\widetilde{J}^r = J^r +e^2\Phi_t(r)H/{4\pi}$
should vanish at the horizon, which determines the value of the charge flux as
\be
D_o = -\frac{e^2}{2\pi}\Phi_t(r_+) = \frac{e^2Q}{2\pi}\ln\frac{r_+}{l} \, .
\label{vcf}
\ee
This shows that, in order to cancel a gauge anomaly at the horizon, a gauge current flux given by Eq. (\ref{vcf})
must be introduced in the anomalous point of view. In fact, this compensating flux is exactly equal to the
electric current flux of Hawking radiation whose Hawking spectrum is given by the Planckian distribution
with a chemical potential for an electric charge of the field radiated from the black hole.

Now, we concentrate on studying the total flux of the energy momentum tensor and gravitational anomaly at the
horizon. Apart from gauge symmetry, the effective two-dimensional theory for each partial wave exhibits
general coordinate symmetry. When the effective theory is formulated outside the black hole horizon to
exclude the offending modes at the horizon, it becomes anomalous with respect to general coordinate symmetry.
To restore general coordinate covariance at the quantum level, one must introduce an energy momentum tensor
flux to cancel gravitational anomaly at the horizon. Similar to the analysis in Sec. \ref{sbh}, in the region
$r_+ +\epsilon \leq r$, since there is an effective background gauge potential, the energy momentum tensor
satisfies the modified conservation equation $\partial_r T_{(o)t}^r = J_{(o)}^r\partial_r \Phi_t(r)$. In the
other region $(r_+ \leq r \leq r_+ +\epsilon)$, when neglecting the classically irrelevant ingoing modes, the
effective two-dimensional chiral theory contains gravitational anomaly at the horizon, and the energy momentum
tensor should satisfy the anomalous equation $\partial_r T_{(H)t}^r = J_{(H)}^r\partial_r\Phi_t(r)
+\Phi_t(r)\partial_r J_{(H)}^r +\partial_rN_t^r(r)$, where the second and third terms come, respectively,
from gauge and gravitational anomalies at the horizon. Under the diffeomorphism transformation, the variation
of the effective action becomes
\bea
-\delta_\lambda W
&=& \int dtdr \lambda^t\Big\{D_o\partial_r\Phi_t(r)
+\partial_r\Big[\frac{e^2}{4\pi}\Phi_t^2(r)H +N_t^r(r)H\Big] \nn \\
&& +\big[T_{(o)t}^r -T_{(H)t}^r +\frac{e^2}{4\pi}\Phi_t^2(r)
+N_t^r(r)\big]\delta(r -r_+ -\epsilon)\Big\} \, ,
\eea
where $N_t^r(r) = \big(f_{,r}^2 +f_{,rr}f\big)/(192\pi)$. In the above derivation, the effective action is
obtained from integrating the outgoing modes. If the ingoing modes are incorporated into the action, the
second term is cancelled by its quantum effect whose contributions are given by $-\big[e^2\Phi_t^2(r)/(4\pi)
+N_t^r(r)\big]H(r)$. The first term is the classical effect of the background electric field for constant current
flow, and the third one is nullified by demanding the effective action to be covariant at the horizon under
the diffeomorphism transformation. We thus have
\be
g_o = g_H+\frac{e^2}{4\pi}\Phi_t^2(r_+) -N_t^r(r_+) \, .
\ee
where $g_o = T_{(o)t}^r -D_o\Phi_t(r)$ is an integration constant, and
\be
g_H = T_{(H)t}^r -\int_{r_+}^r dr \partial_r \big[D_o\Phi_t(r)
+\frac{e^2}{4\pi}\Phi_t^2(r) +N_t^r(r)\big] \, , \nn
\ee
is the value of the energy momentum tensor flux at the horizon. Taking the regularity condition as expressed by
Eq. (\ref{temp}) at the horizon, the compensating flux of the energy momentum tensor reads
\be
g_o = \frac{e^2}{4\pi}\Phi_t^2(r_+) +\frac{\pi}{12}T_+^2 \, ,
\label{emt1}
\ee
where
\be
T_+ = \frac{\kappa}{2\pi}=\frac{2r_+^3 -l^4Q^2}{4\pi l^2r_+^2},
\ee
is the Hawking temperature of the black hole. To restore the general coordinate symmetry of the effective
action, the compensating flux of the energy momentum tensor must take the form of Eq. (\ref{emt1}). At the
quantum level, this flux corresponds to the energy momentum tensor flux of Hawking radiation whose
Hawking spectrum is given by the Planckian distribution with a chemical potential for an electric
charge of the field radiated from the black hole.

Now we turn to derive the Hawking flux of the black hole. Since the Hawking spectrum for fermions of
the charged BTZ black hole is given by the Planckian distribution with an electric chemical potential,
specifically expressed as $N_{\pm e}(\omega) = 1/\big\{\exp[\frac{\omega\pm e\Phi_t(r_+)}{T_+}]
+1\big\}$, the electric current and energy momentum tensor fluxes of Hawking radiation are then given by
\bea
F_Q &=& e\int_0^\infty\frac{1}{2\pi}\Big[N_e(\omega) -N_{-e}(\omega)\Big]d\omega
= \frac{e^2Q}{2\pi}\ln\frac{r_+}{l} \nn \\
F_{H} &=& \int_0^\infty\frac{\omega}{2\pi}\Big[N_e(\omega) +N_{-e}(\omega)\Big]d\omega
= \frac{e^2}{4\pi}\Phi_t^2(r_+) +\frac{\pi}{12}T_+^2 \, .
\label{emt2}
\eea
Comparing Eqs. (\ref{vcf}) and (\ref{emt1}), derived from the conditions of gauge and gravitational anomaly
cancellations and the regularity requirement at the horizon, with Eq. (\ref{emt2}), we easily find that the
fluxes of Hawking radiation from the charged BTZ black hole are capable of cancelling gauge and gravitational
anomalies at the horizon, and of restoring gauge invariance and general coordinate covariance at the quantum
level.

%%%%%%%%%%%%%%%%%%%%%%%%%%%%%%%%%%%%%%%%%%%%%%%%%%%%%%%%%%%%%
\section{Concluding remarks}\label{core}
%%%%%%%%%%%%%%%%%%%%%%%%%%%%%%%%%%%%%%%%%%%%%%%%%%%%%%%%%%%%%

In this Letter, we have investigated Hawking radiation of the $(2+1)$-dimensional BTZ black holes from the
anomalous point of view. As an anomaly often takes place in the spacetime with dimensions $4n+2$ ($n$ is an
integer), we have to first reduce the $(2+1)$-dimensional theory to the two-dimensional effective theory
by a dimensional reduction technique near the horizon. For the charged $(2+1)$-dimensional BTZ black hole,
the scalar field can be effectively described by an infinite collection of two-dimensional quantum field
with a gauge potential, whose gauge charge is the electric charge $e$. As for the rotating $(2+1)$-dimensional
BTZ black hole, the effective theory can be described by the complex scalar field in a two-dimensional charged
black hole, but the $U(1)$ gauge charge $m$ is now referred to by an azimuthal quantum number. However, if
the rotating BTZ black hole is described in the dragging system, each partial wave for the effective theory
also behaves like an independent two-dimensional scalar field, but without a $U(1)$ gauge potential.
Subsequently, in the effective two-dimensional reduction we formulate the effective field theory outside the
horizon to integrate out the horizon-skimming modes, making it chiral and each partial wave exhibits gauge
and gravitational anomalies near the horizon. To demand gauge invariance and general coordinate covariance
at the quantum level, the compensating gauge current and energy momentum tensor fluxes are equal to those
of Hawking radiation.

It should be noted that an observer at rest in the dragging coordinate system, which behaves like a locally
non-rotating coordinate system, would not observe a U(1) gauge current (angular momentum) flux arising
from the rotation of the black hole since he is co-rotating with the black hole. The absence of angular
momentum in the dragging coordinate system is compensated by the frame-dragging effect; this does not
contradict the fact that the flux of angular momentum can be derived by using the Robinson-Wilczek's
method in the Boyer-Lindquist coordinate system. Since the dragging coordinate system is related to the
Boyer-Lindquist coordinate system by the transformation: $\psi = \phi -\frac{J}{2r^2}t$, and $\xi = t$,
so the energy $\omega$ carried by the observer in the dragging coordinates is related by $\omega =
\omega' -m\Omega_+$ to the energy $\omega'$ measured at infinity. If we restore the Boyer-Lindquist
coordinate system from the dragging coordinate system, it is easy to calculate the flux of the angular
momentum.

\section*{Acknowledgments}

This work was partially supported by the Natural Science Foundation of China under Grant Nos. 10675051,
10635020, 70571027 and a grant by the Ministry of Education of China under Grant No. 306022. S.Q.-Wu was
also supported in part by a starting fund from Central China Normal University.

\def\JHEP{J. High Energy Phys. \,}


\begin{thebibliography}{99}

\bibitem{SWH}
S. Hawking, Commun. Math. Phys. \textbf{43} (1975) 199.
% Particle Creation By Black Holes.

\bibitem{JWC}
G. Gibbons, S. Hawking, Phys. Rev. D \textbf{15} (1977) 2752.

\bibitem{PW}
M. Parikh, F. Wilczek, Phys. Rev. Lett. \textbf{85} (2000) 5042;
% Hawking radiation as tunneling, hep-th/9907001.

S. Hemming, E. Keski-Vakkuri, Phys. Rev. D \textbf{64} (2001) 044006;

A.J.M. Medved, Phys. Rev. D \textbf{66} (2002) 124009;

E.C.Vagenas, Phys. Lett. B \textbf{559} (2003) 65;

Q.Q. Jiang, S.Q. Wu, Phys. Lett. B \textbf{635} (2006) 151;

Q.Q. Jiang, S.Q. Wu, X. Cai, Phys. Rev. D \textbf{73} (2006) 064003; \textbf{73} (2006) 06992 (E);

J.Y. Zhang, Z. Zhao, J. High Energy Phys. \textbf{0510} (2005) 055;
Phys. Lett. B \textbf{618} (2005) 14.

\bibitem{CF}
S. Christensen, S. Fulling, Phys. Rev. D \textbf{15} (1977) 2088.
% Trace Anomalies And The Hawking Effect.

\bibitem{RW}
S.P. Robinson, F. Wilczek, Phys. Rev. Lett. \textbf{95} (2005) 011303.
% Relationship between Hawking Radiation and Gravitational Anomalies, gr-qc/0502074.

\bibitem{IUW1}
S. Iso, H. Umetsu, F. Wilczek, Phys. Rev. Lett. \textbf{96} (2006) 151302.
% Hawking Radiation from Charged Black Holes via Gauge and Gravitational Anomalies, hep-th/0602146.

\bibitem{IUW2}
S. Iso, H. Umetsu, F. Wilczek, Phys. Rev. D \textbf{74} (2006) 044017;
% Anomalies, Hawking Radiations and Regularity in Rotating Black Holes, hep-th/0606018.

K. Murata, J. Soda, Phys. Rev. D \textbf{74} (2006) 044018;
% Hawking Radiation from Rotating Black Holes and Gravitational Anomalies, hep-th/0606069.

E.C. Vagenas, S. Das, \JHEP \textbf{0610} (2006) 025;
% Gravitational Anomalies, Hawking Radiation, and Spherically Symmetric Black Holes, hep-th/0606077.

M.R. Setare, Eur. Phys. J. C \textbf{49} (2006) 865.
% Gauge and Gravitational Anomalies and Hawking Radiation of Rotating BTZ Black Holes,

\bibitem{MV}
Z.B. Xu, B. Chen, Phys. Rev. D \textbf{75} (2007) 024041;

S. Iso, T. Morita, H. Umetsu, \JHEP \textbf{0704} (2007) 068;
Phys. Rev. D \textbf{75} (in press), hep-th/0701272;

Q.Q. Jiang, S.Q. Wu, Phys. Lett. B \textbf{647} (2007) 200;

Q.Q. Jiang, S.Q. Wu, X. Cai, Phys. Rev. D \textbf{75} (2007) 064029;

K. Xiao, W.B. Liu, H.B. Zhang, Phys. Lett. B \textbf{647} (2007) 482.

\bibitem{recent}
H. Shin, W. Kim, arXiv: 0705.0265 [hep-th];

J.J. Peng, S.Q. Wu, arXiv: 0705.1225 [hep-th];

Q.Q. Jiang, arXiv: 0705.2068 [hep-th];

S. Das, S.P. Robinson, E.C. Vagenas, arXiv: 0705.2233 [hep-th];

Bin Chen, Wei He, arXiv: 0705.2984 [gr-qc];

U. Miyamoto, K. Murata, arXiv: 0705.3150 [hep-th].

\bibitem{GKP}
S.S. Gubser, I. R. Klebanov, A. W. Peet, Phys. Rev. D \textbf{54} (1996) 3915.
%¡°Entropy And Temperature Of Black 3-Branes,¡±

\bibitem{LE}
L. Alvarez-Gaume, E. Witten, Nucl. Phys. B \textbf{234} (1984) 269;
% Gravitational anomalies.

R. Bertlmann, E. Kohlprath, Ann. Phys. (N.Y.) \textbf{288} (2001) 137.

\bibitem{HA}
P. Kraus, F. Larsen, \JHEP \textbf{0601} (2006) 022;
% Holographic gravitational anomalies, hep-th/0508218.

S.N. Solodukhin, \JHEP \textbf{0607} (2006) 003;
% Holographic description of gravitational anomalies, hep-th/0512216.
Phys. Rev. D \textbf{74} (2006) 024015.
% Holography with a gravitational Chern-Simons term, hep-th/0509148.

\bibitem{BTZ}
M. Ba\~{n}ados, C. Teitelboim, J. Zanelli, Phys. Rev. Lett. \textbf{69} (1992) 1849;
% Black hole in three-dimensional spacetime, hep-th/9204099.

M. Ba\~{n}ados, M. Henneaux, C. Teitelboim, J. Zanelli, Phys. Rev. D \textbf{48} (1993) 1506;
% Geometry of the 2 + 1 black hole, gr-qc/9302012.

C. Martinez, C. Teitelboim, J. Zanelli, Phys. Rev. D \textbf{61} (2000) 104013;

G. Clement, Class. Quant. Grav. \textbf{10} (1993) L49; Phys. Lett. B \textbf{367} (1996) 70.

\bibitem{FL}
S. Wang, S.Q. Wu, F. Xie, L. Dan, Chin. Phys. Lett. \textbf{23} (2006) 1096.
% The first laws of thermodynamics of the (2+1)-dimensional BTZ black holes and Kerr-de Sitter
% spacetimes, hep-th/0601147.

\end{thebibliography}
\end{document}